\documentclass[aps,prl,twocolumn,showpacs,superscriptaddress]{revtex4}

\usepackage{graphicx}
\begin{document}

%\preprint{12}

\title{Density domains of a photo-excited electron gas on liquid helium}
\author{Yu.~P. Monarkha}
\email[E-mail: ]{monarkha@ilt.kharkov.ua}
\affiliation{Institute for Low Temperature Physics and Engineering, 47 Prospekt Nauky,
61103, Kharkov, Ukraine}

\begin{abstract}
The Coulombic effect on the stability range of the photo-excited electron gas on
liquid helium is shown to favor formation of domains of different densities.
Domains appear to eliminate or greatly reduce
regions with negative conductivity. An analysis of the density domain structure allows
explaining remarkable observations reported recently for the photo-excited electron gas.
\end{abstract}


%%73.40.-c    Electronic transport in interface structures
%%73.20.-r    Electron states at surfaces and interfaces
%%78.70.Gq    Microwave and radio-frequency interactions
%%78.67.-n    Optical properties of low-dimensional, mesoscopic,
%%             and nanoscale materials and structures
%%73.25.+i    Surface conductivity and carrier phenomena

%%\keywords{2D electron gas, microwave radiation, negative conductivity, zero resistance states}

\maketitle

Properties of a 2D electron gas exposed to a perpendicular magnetic field $B$
and irradiated with microwaves (MW) recently became a subject of intense
studies because of new interesting discoveries. In GaAs/AlGaAs
heterostructures, novel $1/B$-periodic MW-induced resistance oscillations
(MIRO)~\cite{ZudSim-2001} and zero-resistance states (ZRS)~\cite{ManSme-2002,ZudDu-2003}
were observed, when the MW frequency $\omega $ was larger than the cyclotron
frequency $\omega _{c}=eB/Mc$. The ZRS appears in the vicinity of a resistance minimum
near $\omega /\omega _{c}=m+1/4$, where $m$ is an integer. These
observations have attracted much interest, and a number of theoretical
mechanisms were proposed to explain
MIRO~\cite{DurSac-2003,RyzSur-2003,DmiVav-2005,Mik-2011} and
ZRS~\cite{AndAleMil-2003,FinHalp-2009}.

Similar $1/B$-periodic oscillations of the dc magnetoconductivity $\sigma _{xx}$
and the ZRS were observed
in the 2D electron system formed on the surface of
liquid helium~\cite{KonKon-2009,KonKon-2010}, when the system was tuned to
the resonance with the MW frequency, $\Delta _{2}-\Delta _{1}=\hbar \omega $
(here $\Delta _{l}$ is the energy of surface subbands, $l=1,2,...$ ).
Therefore, the period of these oscillations was actually governed by the
ratio $\omega _{2,1}/\omega _{c}$ where $\omega _{2,1}=\left( \Delta
_{2}-\Delta _{1}\right) /\hbar $. The theory of this effect~\cite{Mon-2011}
is based on nonequilibrium filling of the second
surface subband, $N_{2}>N_{1}\exp (-\hbar \omega _{2,1}/T_{e})$, which
triggers intersubband electron scattering against or along the in-plane dc
electric field $E_{\mathrm{dc}}$ depending on the sign of $\omega
_{2,1}/\omega _{c}-m$. At strong enough power, the sign-changing
contribution to the effective collision frequency can make $\sigma _{xx}<0$,
which by itself suffices to explain the existence of the
ZRS~\cite{AndAleMil-2003}. For surface electrons (SEs) on liquid helium, minima of $%
\sigma _{xx}$ are not strictly fixed to the condition $\omega _{2,1}/\omega
_{c}=m+1/4$. Their positions depend on the strength of Coulomb interaction
between electrons and on the number $m$~\cite{Mon-2012,KonMonKon-2013}.

A state of an electron system with $\sigma _{xx}<0$ is
unstable~\cite{AndAleMil-2003}, and usually a certain local current density $j_{0}$ is
necessary to reach the stable state where $\sigma _{xx}\left( j_{0}\right)
=0 $. This predicts~\cite{AndAleMil-2003} the existence of current domains,
where electrons move in opposite directions.
Such a model of the ZRS is well
applicable for semiconductor electrons. SEs on liquid helium represent a
highly correlated system, where the average Coulomb interaction potential
energy per an electron $U_{C}$ is much larger than the average kinetic
energy. In such a system, current domains
are unlikely due to the strong mutual friction of currents in the domain
wall. Therefore, the internal structure of the ZRS of SEs on liquid helium
is under the question. Experimental observation of an ultra-strong
photovoltaic effect~\cite{KonCheKon-2012} which emerges in the regime
$\sigma _{xx}\rightarrow 0$ gives an
important insight into this problem.
This effect is characterized by a strong displacement of electrons against the
confining force of Corbino electrodes placed below the liquid helium surface
(electrons are displaced from the center of
the plane towards the edge). Remarkably, the redistributed charge exhibits
spontaneously generated oscillations in the audio-frequency
range~\cite{KonWatKon-2013}.

Recently a striking example of irradiation-induced self-organization was
observed~\cite{CheWatKon-2015} in a coupled system of two electron gases of
different densities. The 2D electron gases were formed on the surface of
liquid $^{4}\mathrm{He}$ above the central Corbino electrode (with density
$n_{e}$) and above the guard-ring electrode (with density $n_{g}$). At a
fixed total number of electrons $N_{e}$, the ratio $n_{e}/n_{g}$ was varied
by changing the potential $V_{\mathrm{g}}$ applied to the guard electrode.
In the presence of resonant MW radiation, under the magnetic field fixed to the ZRS
condition, $\omega _{2,1}/\omega _{c}\left( B\right) =6.25$, the inner 2D
electron gas enters an incompressible state with an electron density $%
n_{e}=n_{c}\simeq 3.4\cdot 10^{6}\,\mathrm{cm}^{-2}$ independent of $N_{e}$
and of the potential applied to the guard electrode for
a wide range of parameters.

Surprising new discoveries reported for SEs in the regime of vanishing $%
\sigma _{xx}$ (the photovoltaic effect, self-generated oscillations, and the
incompressible state) have no convincing explanations and require additional
theoretical studies of the nature of the ZRS in a highly correlated electron
system. In this work, we report
results of theoretical study of Coulombic effects on the stability
range of the electron system under conditions of
experiments~\cite{KonCheKon-2012,KonWatKon-2013,CheWatKon-2015}.
We found that for a magnetic field fixed near a magnetoconductivity minimum,
usually there are two electron densities
$n_{\mathrm{H}}$ and $n_{\mathrm{L}}$ (here $n_{\mathrm{H}}>n_{\mathrm{L}}$)
which restrict the unstable region where $\sigma _{xx}<0$.
Values of $n_{\mathrm{H}}$ and $n_{\mathrm{L}}$ are determined only by $B$
and the MW power. Such an unusual dependence of $\sigma _{xx}$ on
electron density allows us to formulate the density domain structure of the
ZRS of SEs on liquid helium which explains recent
observations~\cite{KonCheKon-2012,KonWatKon-2013,CheWatKon-2015}.

It is instructive to consider the dependence of $\sigma _{xx}\left( B\right)
$ in the vicinity of a conductivity minimum using the microscopic mechanism of
magneto-oscillations~\cite{Mon-2012} applicable for electrons with $U_{C}\gg
T $. At low temperatures, the momentum relaxation of SEs is determined by
electron interaction with capillary wave quanta (ripplons). New features of
the theory follow from the structure of the average probability of
intersubband scattering from $l=2$ to $l=1$ accompanied by the momentum
exchange $\hbar \mathbf{q}$ due to ripplon destruction and creation $\nu
_{2\rightarrow 1}\left( \mathbf{q}\right) $, which depends on the average
electron velocity $\mathbf{V}$. Under the conditions $V_{y}\simeq -V_{H}$
and $\left\vert V_{x}\right\vert \ll V_{H}$ (here $V_{H}=cE_{\mathrm{dc}}/B$%
) this quantity is found as~\cite{Mon-2012}
\begin{equation}
\nu _{2\rightarrow 1}\left( \mathbf{q},V_{H}\right) =2p_{2,1}^{2}\left(
q\right) S_{2,1}\left( q,\omega _{2,1}+q_{y}V_{H}\right) ,  \label{e1}
\end{equation}%
where $p_{2,1}^{2}\left( q\right) $ is defined by electron-ripplon
coupling, and $S_{2,1}\left( q,\Omega \right) $ is a
generalization of the dynamic
structure factor of the 2D electron liquid which takes into account that
Landau levels have different collision broadening $\Gamma _{l}$ for
different subbands. The Eq.~(\ref{e1}) has sharp maxima at
$\omega _{2,1}+q_{y}V_{H}\rightarrow m\omega_{c}$ caused by
the Landau level matching ($m=n^{\prime }-n$).

Electron-ripplon scattering is quasi-elastic because typical ripplon
energies $\hbar \omega _{q}\ll \Gamma _{1}$ for $q\lesssim 1/L_{B}$, where $%
L_{B}=\hbar c/eB$.
% is the magnetic length.
In Eq.~(\ref{e1}), the correction $q_{y}V_{H}$ originates from $eE_{\mathrm{dc}}\left(
X^{\prime }-X\right) =\hbar q_{y}V_{H}$.
When evaluating the decay rate of the
excited subband, $q_{y}V_{H}$ can be neglected. Therefore, $S_{2,1}\left(
q,\omega _{2,1}\right) $ and $\nu _{2\rightarrow 1}\left( \mathbf{q}\right) $
have sharp maxima at $\omega _{2,1}\simeq m\omega _{c}$. For
the Gaussian shape of the Landau level density of states~\cite{Ger-1976},
the maxima are also proportional to a Gaussian $G\left(
w\right) =\exp \left( -w^{2}/\tilde{\gamma}^{2}\right) /\sqrt{\pi }\tilde{%
\gamma}$, where~\cite{Mon-2012}
\begin{equation}
w=\frac{\omega _{2,1}}{\omega _{c}}-m-\frac{\Gamma _{2}^{2}}{4T_{e}\hbar
\omega _{c}}-x_{q}\frac{\Gamma _{C}^{2}}{4T_{e}\hbar \omega _{c}},  \label{e2}
\end{equation}%
\begin{equation}
\tilde{\gamma}=\frac{1}{\hbar \omega _{c}}\sqrt{\frac{\Gamma _{1}^{2}+\Gamma
_{2}^{2}}{2}+x_{q}\Gamma _{C}^{2}},\text{ }x_{q}=\frac{q^{2}L_{B}^{2}}{2},
\label{e3}
\end{equation}%
$T_{e}$ is the electron temperature, and $\sqrt{x_{q}}\Gamma _{C}$ is an
additional broadening of the dynamic structure factor induced by Coulomb
interaction. Considering the fluctuational electric
field~\cite{DykKha-1979} $E_{f}$, acting on an electron, as a quasi-uniform field
yields~\cite{MonKon-book-2004} $\Gamma _{C}=$ $\sqrt{2}E_{f}^{\left( 0\right) }L_{B}$, where $%
E_{f}^{\left( 0\right) }\simeq 3\sqrt{T_{e}}n_{s}^{3/4}$ and $n_{s}$ is the
SE density. In Eq.~(\ref{e2}), the third term is very small because $\Gamma
_{l}\ll T_{e}$ and $\Gamma _{l}\ll \hbar \omega _{c}$. The last term can be
substantial at high electron densities or at large level matching numbers $m$
when the average value of the parameter $x_{q}$ increases. For conditions of
experiments~\cite{KonCheKon-2012,KonWatKon-2013,CheWatKon-2015}, it is also
small.

Intersubband scattering $2\rightarrow 1$ is accompanied by the momentum
exchange $\hbar \mathbf{q}$, whose average value can be obtained expanding
Eq.(\ref{e1}) in $q_{y}V_{H}$. Thus, the corresponding contribution into the
electron momentum relaxation rate is proportional to the derivative of the
Gaussian $G^{\prime }\left( w\right) =-2\tilde{\gamma}^{-2}wG\left( w\right)
$, and it changes sign at $\omega _{2,1}/\omega _{c}=m$. This determines the
shape of magnetoconductivity oscillations and allows to estimate positions
of minima. If we neglect overlapping of the sign-changing terms in the sum
over the all $m$, then positions of minima are given by
\begin{equation}
\omega _{2,1}/\omega _{c}-m=\tilde{\gamma}/\sqrt{2}+x_{q}\frac{\Gamma
_{C}^{2}}{4T_{e}\hbar \omega _{c}}.  \label{e4}
\end{equation}%
The deviation from the level
matching point increases monotonically with electron density due to the
Coulomb broadening $\Gamma _{C}$ entering also $\tilde{\gamma}$. At large $m$%
, when the sign-changing terms become strongly overlapping, the positions of
minima are asymptotically given by%
\begin{equation}
\frac{\omega _{2,1}}{\omega _{c}}-m=\frac{1}{4}+x_{q}\frac{\Gamma _{C}^{2}}{%
4T_{e}\hbar \omega _{c}},  \label{e5}
\end{equation}%
where we can use the estimation $x_{q}\sim m$.

The analysis given above indicates that a position of a conductivity minimum
generally depends strongly on $n_{s}$: an increase of $n_{s}$
lifts up the position of the minimum and shifts it right along the $\omega
_{2,1}/\omega _{c}$-axis, as follows from Eqs.~(\ref{e4}) and (\ref{e5}).
This means that changing electron density itself can eliminate the unstable state
with $\sigma _{xx}<0$. It should be noted that experimental data show
somewhat larger shifts of $\sigma _{xx}$ minima than those given by the
theory because it disregards heating of SEs. Still, the theory describes
well the nontrivial dependence of conductivity extrema on $n_{s}$ and
$m$~\cite{KonMonKon-2013}. Therefore, we shall use
the approximation $T_{e}=T$ to obtain stability range of the electron system
under intersubband excitation keeping in mind that decay heating of
SEs can affect our estimations.

\begin{figure}[tbp]
\begin{center}
\includegraphics[width=10.0cm]{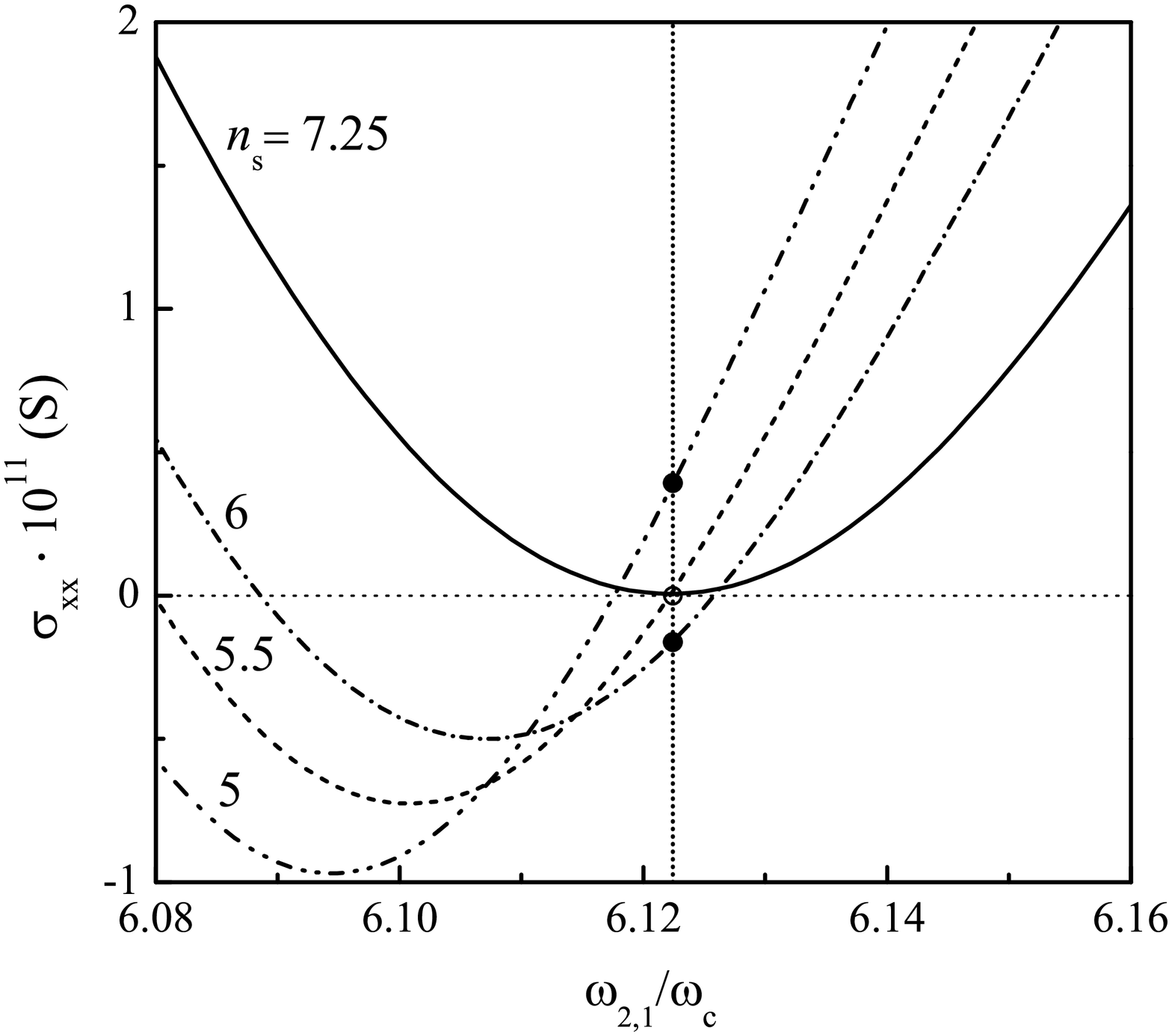}
\end{center}
\caption{ Magnetoconductivity in a dc electric field $E_{dc}$ vs $\omega _{2,1}/\omega_{c}$
calculated for $T=0.3\,\mathrm{K}$ (liquid $^4\mathrm{He}$) and four $n_{s}$.
Values of $n_{s}$ are shown in units of $10^{6}\,\mathrm{cm}^{-2}$.
} \label{f1}
\end{figure}

In the region of interest, the results of numerical evaluations are shown in
Fig.~\ref{f1}. It demonstrates the evolution of the conductivity minimum with
decreasing $n_{s}$ from $7.25\cdot 10^{6}\,\mathrm{cm}^{-2}$
to $5\cdot 10^{6}\,\mathrm{cm}^{-2}$ for a fixed MW power. The vertical line
indicates a fixed magnetic field chosen for our analysis. The cross-points
of this line with curves $\sigma _{xx}\left( \omega _{2,1}/\omega
_{c}\right) $ shown by circles indicate values of $\sigma _{xx}$ for the
given $B$ and $n_{s}$. The open circle indicates the first appearance of the
ZRS. Any shift of the vertical line (chosen value of $\omega _{2,1}/\omega
_{c}$) left or right eliminates this ZRS.

For a lower electron density $n_{s}=6\cdot 10^{6}\,\mathrm{cm}^{-2}$, we have an
unstable state with $\sigma _{xx}<0$ shown by the lower black circle. With
lowering $n_{s}$ curves $\sigma _{xx}\left( \omega _{2,1}/\omega _{c}\right)
$ shift left in accordance with Eqs.~(\ref{e4}) and (\ref{e5}). Eventually,
at $n_{s}=5.5\cdot 10^{6}\,\mathrm{cm}^{-2}$ the corresponding curve $\sigma
_{xx}\left( \omega _{2,1}/\omega _{c}\right) $, again crosses the vertical
line at the point shown by the open circle which indicates the ZRS. Further
reduction in $n_{s}$ leads to a state with $\sigma _{xx}>0$, as shown by the
curve calculated for $n_{s}=5\cdot 10^{6}\,\mathrm{cm}^{-2}$ and by the
higher black circle. Thus, for a fixed magnetic field, the instability
($\sigma _{xx}<0$) appears inside a certain region restricted by two
densities (higher $n_{H}$ and lower $n_{L}$). At $n_{s}=n_{H}$, or at $%
n_{s}=n_{L}$ the system is in the ZRS.

\begin{figure}[tbp]
\begin{center}
\includegraphics[width=10.0cm]{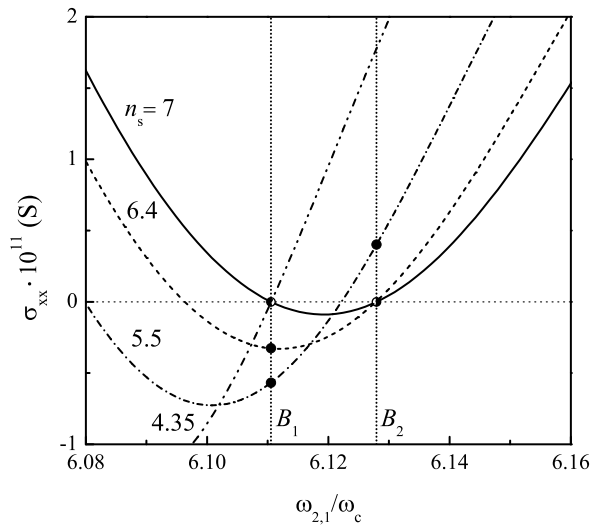}
\end{center}
\caption{ Magnetoconductivity in a dc electric field $E_{dc}$ vs $\omega _{2,1}/\omega_{c}$
calculated for four $n_{s}$. Values of $n_{s}$ are shown in units of $10^{6}\,\mathrm{cm}^{-2}$.
} \label{f2}
\end{figure}

The length of the unstable region $n_{H}-n_{L}$ depends strongly on the
position of the vertical line with regard to the ZRS appeared for the
solid curve of Fig.~\ref{f1}. For example, consider the conditions
illustrated in Fig.~\ref{f2}. Here we chose two magnetic fields $B_{1}$ and $%
B_{2}$ indicated by vertical lines which are located at the left and right
sides with regard to the minima of the solid curve. The crossing points of
the vertical lines with the solid curve
give two ZRS which appear first when decreasing $n_{s}$. One can
see that the range of states with $\sigma _{xx}<0$ strongly differs for $%
B_{1}$ and $B_{2}$. For the right vertical line, it shrinks $%
n_{H}-n_{L}\simeq 0.6\cdot 10^{6}\,\mathrm{cm}^{-2}$ as indicated by the
dashed curve. For intermediate densities, the negative values of $\sigma _{xx}$
are very close to zero, therefore, we didn't show a corresponding
curve in this figure. To the contrary, for the left vertical line, we
found a substantially larger instability range $n_{H}-n_{L}\simeq 1.5\cdot
10^{6}\,\mathrm{cm}^{-2}$. Obviously, $n_{L}$ can be reduced very much
by shifting the vertical line left.
Thus, fixing $B$ and MW power defines two distinct values of electron
density $n_{H}$ and $n_{L}$ where the 2D electron system enters the ZRS.
For intermediate densities $n_{L}<n_{s}<n_{H} $, the system is unstable.

\begin{figure}[t]
\begin{center}
\includegraphics[width=10.0cm]{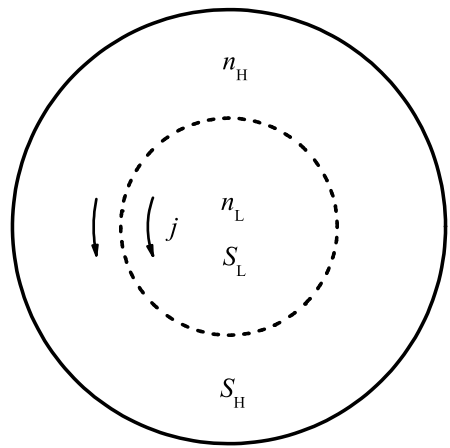}
\end{center}
\caption{ The pattern of the density distribution with a domain wall for the Corbino
geometry.
} \label{f3}
\end{figure}

A negative $\sigma _{xx}$ means that any density fluctuation $\delta n_{s}$
(positive or negative) diffusively grows. The density of growing regions is limited by the conditions:
$n_{s}+\delta n_{s}=n_{H}$ and $n_{s}+\delta n_{s}=n_{L}$. Therefore, the electron system
with $n_{L}<n_{s}<n_{H}$ eventually will be separated into fractions (domains)
with different densities $n_{H}$ and $n_{L}$. For a Corbino geometry,
the simplest stable pattern of the density distribution with a domain wall
is shown in Fig.~\ref{f3}.
Two arrows indicate that in different domains near the domain wall (dashed circle)
local currents flow in the same direction in contrast with
the case of current domains. The position of the domain wall (the areas $%
S_{H}$ and $S_{L}$) corresponding to the initial density $n_{s}$ is
determined by the rule:%
\begin{equation}
\frac{S_{H}}{S_{L}}=\frac{n_{s}-n_{L}}{n_{H}-n_{s}}  \label{e6}
\end{equation}%
We assume that the direction of charge displacement caused by negative
conductivity should be opposite to the direction of the confining force of
Corbino electrodes acting on SEs.

For a sharp domain wall, one can see that the separation of charges shown in
Fig.~\ref{f3} is stable until the resonant MW excitation keeps $\sigma
_{xx}=0 $ in the domains. Assume some electrons by chance have moved from
the area $S_{L}$ to the area $S_{H}$. This makes $\sigma _{xx}>0$ in both
areas and the same amount of electrons will return back due to the inner
electric field induced by charge separation. In a contrary case, when some
electrons have moved from the area $S_{H}$ to the area $S_{L}$, the
magnetoconductivity in both domains became negative, and the same amount of
electrons will return back moving in an uphill direction with regard to the
local dc electric field.
Therefore, under the condition of Eq.~(\ref{e6}) the
domain structure is in a dynamic equilibrium.

Formation of the domain structure caused by negative conductivity
effects explains electron displacement
towards the edge observed under resonant MW radiation~\cite{KonCheKon-2012}.
According to our estimations the displaced fraction of electrons can be of
the order of $N_{e}$ which agrees with observations.
In Ref.~\onlinecite{KonCheKon-2012}, the initial electron
density (in the dark) was about $1.4\cdot 10^{6}\,\mathrm{cm}^{-2}$ which is
smaller than the value of the lower critical density $n_{L}$ defined in Figs.%
~\ref{f1} and \ref{f2}. It should be noted that the calculations
didn't take into account electron heating. For heated electrons, according
to Eqs.~(\ref{e2}) and (\ref{e3}), the same many-electron effect is produced
by an electron density lower by the factor $(T/T_{e})^{2/3}$ than it is for $%
T_{e}=T$. Assuming $T_{e}\approx 2\,\mathrm{K}$, we obtain the factor $0.2$
which makes $n_{L}$ less than the density used in the
experiment~\cite{KonCheKon-2012}.

Regarding self-generated oscillations observed in Ref.~\onlinecite{KonWatKon-2013},
first we note that the frequency of interior magnetoplasmons has a gap $\omega _{c}$, and they cannot
be a reason for audio-frequency oscillations.
SEs on liquid helium have well-defined
audio-frequency excitations: edge magnetoplasmons (EMP)~\cite{MasDahFet-1985,GlaAndDev-1985}
and inter-edge magnetoplasmons (IEMP)~\cite{SomSteHei-1995}. They are localized near
the edge of a 2D electron system (EMP) or near internal boundary of two
contacting regions with different densities (IEMP). Under a strong magnetic
field, the spectrum of the IEMP is gapless, and its frequency decreases with
$B$~\cite{MikVol-1992}%
\begin{equation}
\omega _{iemp}=2q_{y}\left( \sigma _{yx}^{H}-\sigma _{yx}^{L}\right) \left(
\ln \frac{1}{\left\vert q_{y}\right\vert b}+C\right) ,\text{ }  \label{e8}
\end{equation}%
where $\sigma _{yx}^{H/L}\sim n_{H/L}/B$, the constant $C$ depends on
details of the density profile, $b$ is the width of the transition layer,
and $q_{y}$ is the wavevector component along the boundary.
If the inter-edge profile for the domain structure
of Fig.~\ref{f3} is sufficiently smooth, a negative dc conductivity can be
ascribed to a narrow strip of the domain wall. It is quite obvious that this
leads to a negative damping and to self-generation of the IEMP. The same argument is
applicable to the EMP if the net conductivity within the edge
profile is negative. Thus, experimental observation of self-generated
audio-frequency oscillations~\cite{KonWatKon-2013} can be considered as
a convincing evidence for negative conductivity coexisting with the ZRS of
surrounding domains.

\begin{figure}[tbp]
\begin{center}
\includegraphics[width=10.0cm]{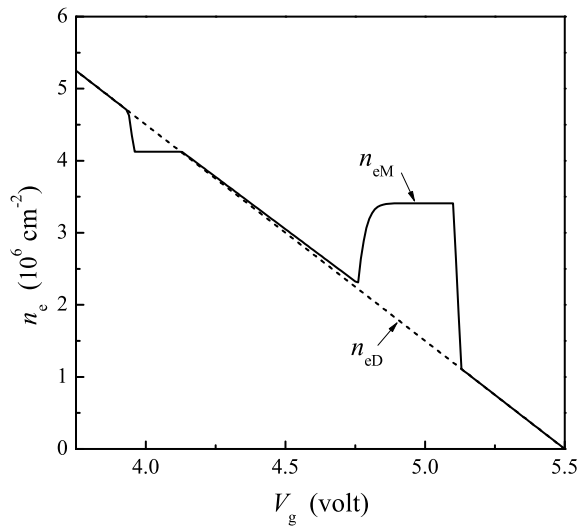}
\end{center}
\caption{ A schematic plot showing how, typically, the electron density, $n_{e}$ , varies with the
guard voltage, $V_{g}$ , in the experiment~\cite{CheWatKon-2015}: the dark case (dashed), irradiated
SEs (solid).
} \label{f4}
\end{figure}

The appearance of two critical electron
densities $n_{H}$ and $n_{L}$ restricting the instability range provides also
an insight into the nature of the incompressible state observed for the
system of two coupled 2D electron gases~\cite{CheWatKon-2015}. In this
experiment, density domains were created artificially by applying different
potentials to the guard ($V_{g}$) and central ($V_{e}$) electrodes of
Corbino geometry. SEs were redistributed between these domains with areal
densities $n_{g}$ and $n_{e}$ by varying $V_{g}$, so that at large $V_{g}$
nearly all SEs were located above the guard-ring electrode ($%
n_{g}=N_{e}/S_{g}$ , $n_{e}=0$). In the opposite limit, SEs were
concentrated above the central electrode: $n_{g}=0$ and $n_{e}=N_{e}/S_{e}$.
Here $S_{e}$ and $S_{g}$ are the areas of the central and guard electrodes
respectively. The experimental dependence of $n_{e}\left( V_{g}\right) $ is
shown schematically in Fig.~\ref{f4}. In the dark case (dashed line), electron
compressibility defined
as~\cite{CheWatKon-2015} $\chi =-dn_{e}/dV_{g}$ is nearly constant.
Under MW radiation the dependence $n_{e}\left( V_{g}\right) $
changes drastically, as shown by the
solid line marked $n_{eM}$. Below $5.13\,\mathrm{V}$ there is a sharp (nearly
vertical) increase of $n_{eM}$ up to a value $n_{c}\simeq 3.4\cdot
10^{6}\,\mathrm{cm}^{-2}$. Then, there is a plateau with $\chi =0$.
Remarkably, in the range $4.75<V_{g}<4.9\,\mathrm{V}$, the system
exhibits a negative (!) compressibility: $\chi <0$. At lower $V_{g}$, the $%
n_{eM}\left( V_{g}\right) $ returns to the dependence observed for the dark
case.

The negative compressibility observed means that the 2D
electron gas squeezes when we apply a potential to stretch it, and
the gas expands when we apply a potential to compress it. This agrees with
our understanding of negative conductivity effects. It should be noted that
already from the existence of regions with $\sigma _{xx}<0$ ($\chi <0$) and $%
\sigma _{xx}>0$ ($\chi >0$) it follows that there should be at least one
incompressible state in between them. The puzzling thing is that the
incompressible state is observed in a quite broad region of $V_{g}$.
Taking into account the results obtained here, we can expect anomalies on the dependence
$n_{eM}\left( V_{g}\right)$ near $n_{H}$ and $n_{L}$.
Then, the plateau value $n_{c}$ of Fig.~\ref{f4} can be naturally ascribed to
$n_{H}$, while the density to which the solid line falls down at $V_{g}\simeq
5.13\,\mathrm{V}$ can be ascribed to $n_{L}$ (note that the later density point is
practically independent of $N_{e}$). Thus, we have $n_{H}\simeq 3.4\cdot 10^{6}\,\mathrm{cm}^{-2}$
and $n_{L}\simeq 1.2\cdot 10^{6}\,\mathrm{cm}^{-2}$. The small upper plateau formed at
$V_{g}\sim 4\,\mathrm{V}$ is close to the condition $n_{g}=n_{L}$.

\begin{figure}[tbp]
\begin{center}
\includegraphics[width=9.0cm]{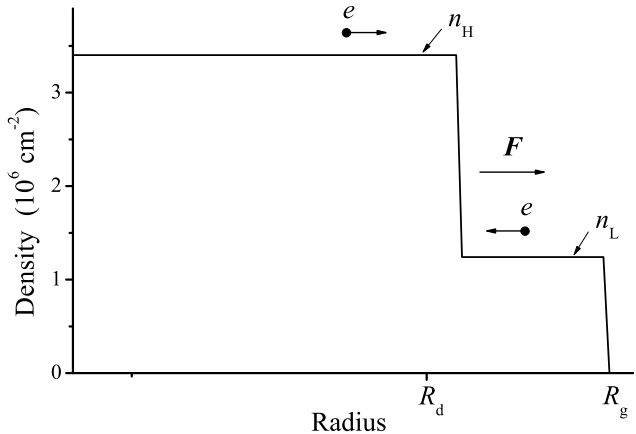}
\end{center}
\caption{ The density domain structure proposed to explain the incompressible state
observed in the experiment~\cite{CheWatKon-2015}.
} \label{f5}
\end{figure}

To explain stability of the state with $n_{e}=n_{H}=n_{c}$ first we note that for
$n_{e}$ a bit lower than $n_{c}$, usually the both densities $n_e$
and $n_g$ enter the unstable region ($n_g$ is somewhat larger than $n_L$).
Therefore, for electrons moving against the confining force, domains with densities $n_{H}$ (center) and $n_{L}$ (edge) should appear to form the ZRS,
as illustrated in Fig.~\ref{f5}. The radius of the central domain $R_{H}$
is usually larger than the radius of the central electrode $R_{d}$. Therefore, the average
density above the central electrode does not change with varying $N_{e}$.
The density distribution shown in Fig.~\ref{f5} is quite stable, and it should not change much with varying
$V_g$. Assume that an electron somehow is displaced from the center to the edge domain. Then, the both regions enter the regime $\sigma_{xx}<0$, and an electron will move in the uphill direction with regard to the total force $\mathbf{F}$ back to restor the ZRS, as shown in Fig.~\ref{f5}. The opposite displacement of an electron makes $\sigma_{xx}>0$ and an electron will move back. Thus, redistribution of electrons between domains is locked out.

According to Eqs.~(\ref{e2}) and (\ref{e3}) the Coulombic effect increases with $m$.
Therefore, at larger $m$ the $n_{H}$ should be lower, which also agrees with observations~\cite{CheWatKon-2015}.

In summary, we have investigated theoretically the influence of Coulomb interaction acting
between electrons on the stability range of the photo-excited electron gas on liquid helium.
The analysis given here allows us to conclude that the zero resistance state of SEs is formed of two
domains of different densities. This conception gives explanations for the photovoltaic effect~\cite{KonCheKon-2012}, self-generated audio-frequency oscillations~\cite{KonWatKon-2013}, and an incompressible state~\cite{CheWatKon-2015} recently observed in experiments.

\end{document}